\documentclass[12pt,preprint]{aastex}

\newcommand{\sect}[1]{\S\,\ref{#1}}

\newcommand{\be}{\begin{displaymath}}
\newcommand{\ee}{\end{displaymath}}
\newcommand{\bea}{\begin{eqnarray}}
\newcommand{\eea}{\end{eqnarray}}

\newcommand\cs{\,cm$^2$\,s$^{-1}$}

\shortauthors{Denissenkov and Pinsonneault}
\shorttitle{Revised Prescription for the Tayler-Spruit Dynamo}

\begin{document}

\title{A REVISED PRESCRIPTION FOR THE TAYLER-SPRUIT DYNAMO: MAGNETIC ANGULAR MOMENTUM
       TRANSPORT IN STARS}

\author{Pavel A. Denissenkov\altaffilmark{1,2}, Marc Pinsonneault\altaffilmark{1}}
\altaffiltext{1}{Department of Astronomy, The Ohio State University, 4055 McPherson Laboratory, 
       140 West 18th Avenue, Columbus, OH 43210; dpa@astronomy.ohio-state.edu, pinsono@astronomy.ohio-state.edu.}
\altaffiltext{2}{On leave from Sobolev Astronomical Institute of St. Petersburg State University,
   Universitetsky Pr. 28, Petrodvorets, 198504 St. Petersburg, Russia.}

\begin{abstract}
Angular momentum transport by internal magnetic fields is an important ingredient 
for stellar interior models. In this paper we critically examine the basic
heuristic assumptions in the
model of the Tayler-Spruit dynamo, which describes how a pinch-type instability of 
a toroidal magnetic field in differentially rotating stellar radiative zones
may result in large-scale fluid motion.
We agree with prior published work both on the existence of the instability and its
nearly horizontal geometry for perturbations.
However, the approximations in the original Acheson
dispersion relation are valid only for small length scales, and we disagree that
the dispersion relation can be extrapolated to horizontal length scales
of order the radius of the star.  We contend that dynamical effects, in particular
angular momentum conservation, limit the maximum horizontal length scale.
We therefore present transport coefficients for chemical
mixing and angular momentum redistribution by magnetic torques that are 
significantly different from previous published values.
The new magnetic viscosity is reduced by 2 to 3 orders of magnitude compared to the old one,
and we find that magnetic angular momentum transport by this mechanism is 
very sensitive to gradients in the mean molecular weight.  The revised coefficients are more
compatible with empirical constraints on the timescale of core-envelope coupling 
in young stars than the previous ones.  However, solar models including only
this mechanism possess a rapidly rotating core, in contradiction with 
helioseismic data.  Previous studies had found strong core-envelope coupling, both for
solar models and for the cores of massive evolved stars.  
We conclude that the Tayler-Spruit mechanism may be important for envelope angular momentum
transport, but that some other process must be responsible for efficient spindown of stellar cores.

\end{abstract}


\section{Introduction}
\label{sec:intro}

Mixing driven by stellar rotation is important for a variety of problems 
in stellar structure and evolution.  Hydrodynamical mechanisms, such as meridional 
circulation and shear instabilities, induce mixing that is consistent with 
observational data (see \citealt{p97,mm00} for reviews 
on rotational mixing in low and high mass stars, respectively).  However, the evolution of a rotating 
star depends on its internal angular momentum distribution, and existing 
theoretical models have difficulty matching empirical constraints on the timescale 
for transport in stellar radiative interiors.  The slow rotation of the solar 
core (\citealt{tea95}) and newly born pulsars (e.g., \citealt{oea05}) indicates that 
angular momentum is transferred more effectively than would be expected from 
hydrodynamical mechanisms alone.  To complicate matters further, solid body rotation is 
also not an appropriate approximation for important phases of stellar evolution.  
Both the spindown of young cluster stars (\citealt{sh87,kea95,bea97,kea97}) and the survival of 
rapid rotation in old horizontal branch stars (\citealt{p83,b03,sp00})
require that internal differential rotation persist for 
timescales of order 20 Myr to 100 Myr. 

There are two other major classes of mechanisms that could be important agents for 
angular momentum transport: magnetic fields and waves (g-modes).  
Both waves and magnetic fields have the property that they generally transmit 
angular momentum much more effectively than they induce mixing.  As a result, 
the efficiency of these processes is most easily tested by their impact on 
angular momentum evolution.  In this paper we focus on magnetic angular momentum 
transport via the Tayler-Spruit mechanism.  

Elaborating upon results of earlier investigations of instabilities of toroidal
magnetic fields in stars conducted by Tayler and others (e.g., \citealt{t73,mt73,mt74,a78,pt85}),
\cite{s99} has proposed a new magnetohydrodynamical mode of angular momentum 
transport in radiative zones of 
differentially rotating stars. Fluid elements experience large-scale horizontal 
displacements caused by an unstable configuration of 
the toroidal magnetic field (one consisting of stacks of rings concentric with the rotation axis). 
Small-scale vertical displacements of fluid elements are coupled to the horizontal motions, 
which can cause both mild mixing and much more effective momentum transport.

Spruit's key idea
is that no initial toroidal magnetic field is actually required
to drive the instability and mixing  because the unstable field configuration
can be generated and maintained by differential rotation itself in a process similar to convective dynamo. 
The dynamo cycle consists of two consecutive steps: first,
a poloidal field $B_r$ is generated by the vertical displacements of the unstable toroidal field; second,
the new poloidal field is stretched into a toroidal field $B_\varphi$ by differential rotation
(for more details, see \sect{sec:heuristic}).

Following \cite{s99,s02}, who considered two opposite limiting cases in which the thermal diffusivity
$K$ is either negligible or dominating,
\cite{mm04} have derived equations for the transport by the Tayler-Spruit dynamo applicable 
to a general case. These equations as well as Spruit's original equations
have already been applied to study the angular momentum transport by magnetic torques,
which are proportional to the product $B_rB_\varphi$, in massive stars
on the main sequence (MS) (\citealt{mm03,mm04,mm05}) and during their entire evolution 
toward the onset of iron-core collapse
(\citealt{hea05,wh05}). While the Geneva group has shown that the Tayler-Spruit dynamo can easily
enforce the solid-body rotation of massive MS stars, Heger and co-workers have established that
appropriate mass-loss rates coupled with the magnetic torques generated by the Tayler-Spruit dynamo
can spin down pulsars to the observed rotation rates.
However, the same angular momentum transport model has failed to explain gamma-ray bursts
(\citealt{wh04}). Finally, \cite{eea05} have recently demonstrated that the Tayler-Spruit dynamo
leads to a nearly flat solar rotation profile consistent with that revealed by the helioseismic data
(e.g. \citealt{cea03}).
The important question of stellar spindown was not directly addressed in this
theoretical work.

In this paper, we derive new transport coefficients
for the Tayler-Spruit dynamo
that have quite different magnitudes and dependences on the $\mu$-gradient 
(\sect{sec:heuristic} and \sect{sec:sun}). 
We review prior work in \sect{sec:heuristic}. 
The diffusion coefficients are obtained as products of the growth rate and 
the characteristic vertical length scale of the instability. 
Spruit (1999) used a local dispersion relation to infer the ratio of vertical to horizontal
length scales. To convert this ratio to an absolute vertical length scale he assigned
the maximum possible horizontal length scale $r$.  We contend that the assumptions in
the dispersion relation break down for large length scales, and that the maximum effective
horizontal length scale cannot be obtained by the method employed by Spruit.
In \sect{sec:arguments}, we give some geometric, energy and dynamical arguments
that support reduced length scales and we present revised diffusivity estimates. 
The original prescription predicts extremely rapid 
core-envelope coupling for low mass stars (in contradiction with observations), 
but also a flat solar rotation curve (in agreement with observations).  
The situation is reversed with our diffusion coefficients; the spindown timescale is 
closer to that inferred from young open cluster stars, but a solar model computed 
with the new equations possesses strong differential rotation in its radiative core, in conflict
with helioseismology. In \sect{sec:sun}, we discuss the observational consequence. We therefore conclude that some other 
magnetic mechanism, or waves, must be responsible for angular momentum coupling 
between deep stellar cores and envelopes.

\section{Revised Heuristic Assumptions and Transport Coefficients}
\label{sec:heuristic}

The diffusion coefficient estimates for the Tayler-Spruit mechanism were derived on a heuristic
level, and the results were confirmed by an appeal to a restricted form of the full
\cite{a78} dispersion relation. We begin this section by summarizing the overall arguments in 
Spruit (1999, 2002), noting the places where we differ with him.
In particular, we claim that the results from the dispersion relation can only be applied
to small horizontal length scales, therefore they cannot be used to justify a characteristic length
scale for horizontal motions of order the radius of the star. In addition, the presence of
curvature terms and variations in physical quantities such as the rotation rate
during the motions will make both our estimates and those of Spruit upper limits.

The basic parameters of the Tayler-Spruit dynamo include the instability's growth rate $\omega$, the
magnetic diffusivity $\eta$, buoyancy (or Brunt-V\"{a}is\"{a}la) frequency $N$, angular velocity $\Omega$ and its
gradient $q=(\partial\ln\Omega/\partial\ln r)$ in a stellar radiative zone.
The vertical displacement $l_r$ of a disturbed fluid element has an upper limit

\bea
l_r^2 < \frac{r^2\omega^2}{N^2}
\label{eq:up}
\eea 
(appropriate expressions for $\omega$ and $N^2$ are provided below).
The displacement is constrained by the requirement that its kinetic energy must exceed 
the work that has to be done against the Archimedes force. 
At the heuristic level, Spruit justified this equation in terms of energy conservation.
He balanced the magnetic energy gain (from motions perpendicular to the rotation axis)
with the work done against the buoyancy force (from motions along the radial direction).
This is an upper limit, as it assumes that all of the magnetic energy can be converted into
vertical motions. He justified this in the dispersion relation by adopting a simplified version
of the full Acheson (1978) equation valid only near the rotation axis.  In this
case, the horizontal and vertical directions correspond to motions perpendicular and
along the rotation axis respectively.  The ratio is maximized at the pole, which is another
way in which this assumption serves as an upper limit rather than a bound.
One can obtain a constraint (equation A21 in Spruit 1999) on the ratio of
the horizontal and vertical length scales. In cylindrical polar coordinates $(w,\varphi,z)$,
used by Acheson and Spruit, this has the general form
$ l_w^2 \omega_{\rm A}^2 > l_z^2 N^2$ for an $m=1$ mode and Spruit's $a=1$.
In the dispersion relation that was used, there is
no bound on the energy gain from horizontal motions; Spruit set
the horizontal length scale to be the radius of the star, which is certainly a maximum.
We will argue in \sect{sec:arguments} that the large scale horizontal fluid motions initiated by the
Tayler instability are restricted by the Coriolis force within bounds of order $l_w\ll r$,
and that the basic assumptions in the dispersion equation, for instance 
those that both $l_w$ and $l_z$ should be much smaller than $r$ (see $\S\,2$ in
\citealt{a78}), imply that this equation cannot be
used to extract information about the absolute maximum horizontal displacement.  
In particular, the
subtle assumption that motions perpendicular to the rotation axis do no work against
the buoyancy force is valid only for small perturbations close to the rotation axis.

There is also a lower limit 

\bea
l_r^2 > \frac{\eta}{\omega},
\label{eq:low}
\eea 
where $\eta$ is the microscopic magnetic diffusivity ($\eta = 7\times 10^{11}\ln\Lambda T^{-3/2}$\,\cs,\ 
where $\ln\Lambda\approx 5$\,--\,10 is the Coulomb logarithm). This bound is determined
by the length scale at which the magnetic diffusion dissolves any new poloidal field faster than it is produced by the instability. 
According to \cite{pt85}, the instability's growth rate takes on a value

\bea
\omega = \omega_{\rm A} \equiv \frac{B_\varphi}{(4\pi\rho)^{1/2}\,r},\ \ \mbox{when}\ \ \Omega\ll\omega_{\rm A},
\label{eq:s1}
\eea
and
\bea
\omega = \frac{\omega_{\rm A}^2}{\Omega},\ \ \mbox{when}\ \ \Omega\gg\omega_{\rm A},
\label{eq:s2}
\eea
where $\omega_{\rm A}$ is the Alfv\'{e}n frequency, $\rho$ and $r$ are the local density and radius.

We are mainly interested in the case of fast rotation, when $\Omega\gg\omega_{\rm A}$.
This condition is satisfied even for the relatively slow solar rotation. 
In this case, the Coriolis force reduces the instability's growth rate considerably
by a factor of $\omega_{\rm A}/\Omega\ll 1$ (\citealt{pt85}). 

In \sect{sec:arguments}, we will argue that one should employ the same value of $\omega$, either (\ref{eq:s1}) or (\ref{eq:s2}),
when estimating the heuristic limits (\ref{eq:up}) and (\ref{eq:low}). 
Below, we present revised diffusivities obtained under this basis.
However, \cite{s99} and, after him, other researchers
(e.g., \citealt{mm04}) have substituted the plain Alfv\'{e}n frequency (\ref{eq:s1}) 
into inequality (\ref{eq:up}) but the reduced growth rate (\ref{eq:s2}) into
inequality (\ref{eq:low}) and all other equations. Having done that, the appropriate heuristic expressions for 
the effective magnetic diffusivity $\eta_{\rm e}\sim l_r^2\,\omega$
and viscosity $\nu_{\rm e}=B_rB_\varphi/(4\pi\rho\, q\,\Omega)$
are obtained quite straightforwardly. 
We therefore repeat the steps in their derivation and note the places where our different
assumptions yield different conclusions.
When deriving them, we will adhere to the following three important
assumptions made by \cite{s99} and \cite{mm04}. 
First, they have considered the case of marginal stability, i.e.
$l_r^2$ has been set equal to its upper limit (\ref{eq:up}). 

From the magnetic induction equations
for $B_r$ and $B_\varphi$ (see below) 
it follows that the joint operation of the Tayler-Spruit instability and differential rotation
can reach a stationary regime, in which the poloidal and toroidal fields maintain each other, only if

\bea
l_r\approx\frac{r\omega}{q\Omega}.
\label{eq:stat}
\eea

Indeed, let us denote $\mathbf{B}=B_r\mathbf{e}_r$ a weak poloidal magnetic field (here, we use the spherical
polar coordinates). If, for the sake of simplicity, we only consider gas motion in the equatorial
plane then its velocity due to rotation is
$\mathbf{v}=\Omega r\mathbf{e}_\varphi$. Let us also neglect the magnetic diffusivity $\eta$ for a while.
Under these assumptions, the magnetic induction equation
$$
\frac{\partial\mathbf{B}}{\partial t} = \nabla\times(\mathbf{v}\times\mathbf{B}) + \eta\nabla^2\mathbf{B}
$$
is reduced to
\bea
\frac{\partial B_\varphi}{\partial t} = \frac{1}{r}\frac{\partial}{\partial r}(\Omega\,r^2B_r)
\label{eq:magind1}
\eea
or, after choosing a frame of reference rotating with the angular velocity $\Omega$,
\bea
\frac{\partial B_\varphi}{\partial t} = r\frac{\partial\Omega}{\partial r}B_r = \Omega\,q B_r,
\label{eq:magind2}
\eea
where $q=(\partial\ln\Omega/\partial\ln r)$. This equation shows how
the differential rotation winds up a toroidal field $B_\varphi\mathbf{e}_\varphi$
by stretching the poloidal field lines around the rotation axis.
However, the azimuthal field is subject to the Tayler-Spruit instability at any field strength (\citealt{pt85,s99}).
This non-axisymmetric pinch-type instability, whose driving force is the magnetic pressure $B_\varphi^2/8\pi$,
causes the concentric field lines to slip sideways a horizontal distance $l_{\rm h}$ accompanied by
a radial displacement $l_r$. From the Acheson equation, Spruit derived the estimate
$l_r/l_{\rm h}\la\omega_{\rm A}/N$.

Like the differential rotation generates the azimuthal magnetic field when operating on the seed poloidal field
(eq. \ref{eq:magind1}), the radial displacement $l_r$ with a characteristic velocity $v_r\approx l_r\,\omega$ produces
a weak poloidal magnetic field $B_r\mathbf{e}_r$ at the expense of the existing toroidal field:
\bea
\frac{\partial B_r}{\partial t} = \frac{1}{r}\frac{\partial}{\partial r}(v_rrB_\varphi)
\approx\frac{v_r}{r}B_\varphi\approx \frac{l_r}{r}\omega B_\varphi.
\label{eq:magind3}
\eea
After that, the differential rotation can operate on the freshly generated poloidal field to wind it up into
a new azimuthal field, and so on. Hence, a coordinated action between the differential rotation and
the radial displacements caused by the Tayler-Spruit instability can work as a dynamo, provided
that some conditions leading to a stationary regime are realized. The stationary regime is
achieved when the magnitude of the poloidal field $B_r$ generated by the instability during its growth time
$\tau\sim\omega^{-1}$, as described by equation (\ref{eq:magind3}), coincides with the magnitude of the field $B_r$
in equation (\ref{eq:magind2}) whose lines can be stretched by the differential rotation, during the same time $\tau$,
into an azimuthal field of the same magnitude $B_\varphi$ as the one that started this dynamo loop.
Quantitatively, these requirements are expressed as
\bea
B_r\approx\tau\frac{\partial B_r}{\partial t}\approx\frac{l_r}{r}B_\varphi\approx\frac{l_r}{r}\tau\frac{\partial B_\varphi}{\partial t} =
\frac{l_r}{r}\tau\Omega qB_r,
\label{eq:req}
\eea
which results in equation (\ref{eq:stat}).

Third, one must take into account the reduction of the stable thermal stratification of the radiative zone
caused by heat losses from surfaces of the disturbed fluid elements during their vertical displacements.
Following \cite{mm04}, for $\eta_{\rm e}\ll K$ the reduced buoyancy frequency can be expressed as
\bea
N^2 = \frac{\eta_{\rm e}/K}{\eta_{\rm e}/K + C}N_T^2 + N_\mu^2,
\label{eq:N2}
\eea
where
\be
N_T^2 = \frac{g\delta}{H_P}(\nabla_{{\rm ad}}-\nabla_{{\rm rad}}),
\ \ \mbox{and}\ \
N_\mu^2 = g\varphi\,\left|\frac{\partial\ln\mu}{\partial r}\right|
\ee
are the $T$- and $\mu$-component of the square of the
buoyancy frequency in the absence of mixing. Here,
$\nabla_{\rm rad}$ and $\nabla_{\rm ad}$ are the radiative
and adiabatic temperature gradients (logarithmic and with respect to
pressure), $H_P$ is the pressure scale-height, $g$ is the local gravity,
$\mu$ is the mean molecular weight, and
\bea
K = \frac{4acT^3}{3\kappa\rho^2C_P}
\eea
is the thermal diffusivity with $\kappa$ and $C_P$ representing the opacity and
the specific heat at constant pressure, respectively. The quantities
$\delta =
-\left(\partial\ln\rho/\partial\ln T\right)_{P,\mu}$ and
$\varphi = \left(\partial\ln\rho/\partial\ln\mu\right)_{P,T}$
are determined by the equation of state. The value of the constant $C$ in equation (\ref{eq:N2})
depends in general on the assumed geometry. Following \cite{mm04}, we adopt $C=2$.

The effective magnetic viscosity can be expressed in terms of $\eta_{\rm e}$:
\bea
\nu_{\rm e} = \left(\frac{r^2\Omega\eta_{\rm e}^2}{q^2}\right)^{1/3}\gg\eta_{\rm e}.
\label{eq:nue}
\eea
The last formula has been derived using magnetic induction equations for $B_r$ and $B_\varphi$ 
(see eqs. \ref{eq:magind2} and \ref{eq:magind3}) in a way similar to that described by \cite{s99}.

To summarize, in general we have 9 independent equations
\bea
l_r = r\frac{\omega_1}{N}, & \omega_{\rm A} = \frac{B_\varphi}{\sqrt{4\pi\rho}\,r}, & l_r = \frac{r\omega_2}{q\Omega}, \nonumber \\
B_r = \frac{l_r}{r}B_\varphi, & N^2 = \frac{\eta_{\rm e}/K}{\eta_{\rm e}/K+C}N_T^2 + N_\mu^2, & \eta_{\rm e} = l_r^2\omega_2, \nonumber \\
\nu_{\rm e} = \frac{B_rB_\varphi}{4\pi\rho q\Omega}, & \omega_1 = \omega_1(\omega_{\rm A},\Omega), & \omega_2 = \omega_2(\omega_{\rm A},\Omega) \nonumber
\eea
for 9 unknown quantities
$$
l_r,\ \omega_{\rm A},\ B_r,\ B_\varphi,\ N,\ \eta_{\rm e},\ \nu_{\rm e},\ \omega_1,\ \mbox{and}\ \omega_2.
$$
For the case of fast rotation, Spruit has taken $\omega_1 = \omega_{\rm A}$ and $\omega_2 = \omega_{\rm A}^2/\Omega$.
In Appendix A to their paper, \cite{mm05} have assumed that $\omega_1 = \omega_2 = \omega_{\rm A}$, which is correct
for the case of slow rotation considered there. Unlike \cite{s99}, we have chosen $\omega_1 = \omega_2 = \omega_{\rm A}^2/\Omega$.
In \sect{sec:arguments}, we use arguments from energetics,
dynamics, and a discussion of the limitations of the usage of the dispersion relation to justify this choice.

Assigning a value of the upper limit (\ref{eq:up}) to $l_r^2$ and comparing it with the expression
for $l_r$ from (\ref{eq:stat}), we obtain $N=q\Omega$. After substituting this
into (\ref{eq:N2}), we find the effective magnetic diffusivity
\bea
\eta_{\rm e} = 2K\frac{\Omega^2 q^2 - N_\mu^2}{N_T^2 + N_\mu^2 - \Omega^2 q^2}.
\label{eq:etae}
\eea

In order to compare our revised transport coefficients for the Tayler-Spruit dynamo
with those used by \cite{mm04}, the latter are summarized below in a slightly modified form:
\bea
\eta_{\rm e} = \alpha\,Ky^3,\ \ \mbox{and}\ \
\nu_{\rm e} = \alpha\,K\frac{N_T^2+N_\mu^2}{\Omega^2q^2}y^2.
\label{eq:mm04}
\eea
Here, $y$ is a solution of the 4th order algebraic equation
\bea
\alpha\,y^4 - \alpha\,y^3 + \beta\,y - 2 = 0,
\label{eq:4th}
\eea
where
\bea
\alpha = r^2\frac{\Omega^7\,q^4}{K(N_T^2+N_\mu^2)^3},\ \ \mbox{and}\ \ \beta = 2\frac{N_\mu^2}{N_T^2+N_\mu^2}
\label{eq:4thcoeff}
\eea
are dimensionless coefficients.

Originally, \cite{s99,s02} only considered the two limiting cases: $N_T\gg N_\mu$ and $N_T\ll N_\mu$.
In the first case, we can neglect the $\beta$-term in equation (\ref{eq:4th}) and re-write it as
$y^3(y-1)\approx 2/\alpha$, where
\bea
\alpha \approx 4.84\times 10^{-2}
\left(\frac{r}{R_\odot}\right)^2\left(\frac{\Omega}{10^{-5}}\right)^7\left(\frac{10^6}{K}\right)
\left(\frac{10^{-3}}{N_T}\right)^6q^4.
\eea
The normalizations in the last equation assume that all quantities are expressed in cgs units.
In the Sun's radiative core (at least, in layers located at some distance from both its center and the bottom of
its convective envelope), $r<R_\odot$, $\Omega<10^{-5}$, $10^5\la K\la 10^7$,
$N_T\ga 10^{-3}$, and $q\ll 1$. Hence, there we have $\alpha \ll 1$, which results in $y\approx (2/\alpha)^{1/4}$.
Using this approximate solution, we estimate
\bea
\eta_{\rm e}\approx 2^{3/4}\alpha^{1/4}K\approx 7.89\times 10^5
\left(\frac{r}{R_\odot}\right)^{1/2}\left(\frac{\Omega}{10^{-5}}\right)^{7/4}\left(\frac{K}{10^6}\right)^{3/4}
\left(\frac{10^{-3}}{N_T}\right)^{3/2}q,\ \ \mbox{cm}^2\,\mbox{s}^{-1}.
\label{eq:oldetae}
\eea
Except for the numerical coefficient and parameter normalizations,
this equation appears to be identical with eq. (43) from \cite{s02} and eq. (23) from \cite{mm04}.

For the opposite case of $N_\mu\gg N_T$, we have $\beta\approx 2$. Therefore, eq. (\ref{eq:4th}) can be re-written as
$(\alpha y^3+2)(y-1)\approx 0$. Since $\alpha$, with $N_T$ replaced by $N_\mu$, still remains a small
quantity, $y\approx 1$ is now a solution. In this case,
\bea
\eta_{\rm e}\approx\alpha K\approx 4.84\times 10^4
\left(\frac{r}{R_\odot}\right)^2\left(\frac{\Omega}{10^{-5}}\right)^7
\left(\frac{10^{-3}}{N_\mu}\right)^6q^4,\ \ \mbox{cm}^2\,\mbox{s}^{-1}.
\eea
This equation coincides with eq. (42) from \cite{s02} and eq. (21) from \cite{mm04}.

In a region where $N_\mu\ll N_T$, our magnetic diffusivity (\ref{eq:etae}) is estimated as
\bea
\eta_{\rm e}\approx 2.00\times 10^2\left(\frac{K}{10^6}\right)
\left(\frac{\Omega}{10^{-5}}\right)^2
\left(\frac{10^{-3}}{N_T}\right)^2q^2,\ \ \mbox{cm}^2\,\mbox{s}^{-1}.
\label{eq:newetae}
\eea
Its values are smaller than those given by the old prescription (\ref{eq:oldetae}) by a factor of
$\sim$\,$10^3$. In the same region, the magnetic field strengths (in Gauss) are
\bea
B_\varphi\approx 9.88\times 10^3\,\rho^{1/2}
\left(\frac{r}{R_\odot}\right)^{2/3}\left(\frac{\Omega}{10^{-5}}\right)^{7/6}\left(\frac{K}{10^6}\right)^{1/6}
\left(\frac{10^{-3}}{N_T}\right)^{1/3}q^{2/3},
\eea
\bea
B_r\approx 0.159\,\rho^{1/2}
\left(\frac{\Omega}{10^{-5}}\right)^{3/2}\left(\frac{K}{10^6}\right)^{1/2}
\left(\frac{10^{-3}}{N_T}\right)q.
\eea

In all three cases considered above (eqs. \ref{eq:oldetae}\,--\,\ref{eq:newetae}),
the effective magnetic viscosity and diffusivity are related
to each other by equation (\ref{eq:nue}), which gives
\bea
\nu_{\rm e} = 3.65\times 10^5\left(\frac{r}{R_\odot}\right)^{2/3}
\left(\frac{\Omega}{10^{-5}}\right)^{1/3}\left(\frac{\eta_{\rm e}}{q}\right)^{2/3}\gg\eta_{\rm e}.
\label{eq:nue2}
\eea

In our case, as has been shown, $N=\Omega\, q$. Interestingly,
the same constraint on the buoyancy frequency  has also been obtained by \cite{mm05} in Appendix to their paper,
where they considered the alternative case of slow rotation, $\Omega\ll\omega_{\rm A}$. 
This coincidence is not surprising
because they have substituted the plain Alfv\'{e}n frequency into the lower limit
(\ref{eq:low}) as well, as prescribed by equation (\ref{eq:s1}) for slow rotation. This also explains why
our revised effective diffusivity (\ref{eq:etae}) coincides with theirs.
However, our expression for the magnetic viscosity (\ref{eq:nue}), which controls the transport of angular momentum,
is different from that derived by \cite{mm05} (they got $\nu_{\rm e}=\eta_{\rm e}$) for slow rotation because
we consider the case of fast rotation with $\omega = \omega_{\rm A}^2/\Omega$ in
all equations. If we had examined the slow rotation case, we would recover Maeder \& Meynet's result in that limit.
Finally, it can easily be shown that our revised equations pass the consistency check demanding that
the rate of magnetic energy production is equal to the rate of the dissipation of
rotational energy by the viscosity $\nu_{\rm e}$.

\section{Arguments Supporting Our Heuristic Assumptions}
\label{sec:arguments}

We begin the justification of our heuristic assumptions by critically
analyzing the validity of the
form of the dispersion relation employed by Spruit (1999)\footnote{In the original version of
our paper (astro-ph/0604045), our mathematical justification
based on a re-analysis of the Acheson dispersion relation
contained subtle flaws. Our conditions (24)
were actually sufficient but not necessary conditions for the development of the Tayler-Spruit instability
(we are grateful to  Tony Piro
for bringing our attention to this). Although now we agree with the estimate (A21)
obtained by Spruit (1999), we disagree with his implementation of it, particularly with
Spruit's choice of the horizontal length scale.}.
In this initial section we demonstrate that the approximations in the
dispersion relation at the rotation axis are valid only
for small horizontal length scales and cannot be extended to a length scale
comparable to that of the star, as was
assumed by Spruit.  
The dispersion relation sets only the ratio of length scales, and can only rigorously be used
for perturbations. Our strongest result is therefore that the coefficients presented by Spruit
must be treated as upper bounds.

We then follow up with a discussion of the dynamics to obtain other information about behavior 
for larger displacements. 
We argue that rotation
limits the maximum horizontal length scale with the same damping factor as
present for the small scales where diffusion
becomes important.  The energy constraints from the magnetic induction
equations are then considered, and we conclude with some
general comments. 

\subsection{Geometric Arguments}
\label{sec:geometry}

\begin{figure}
\plotone{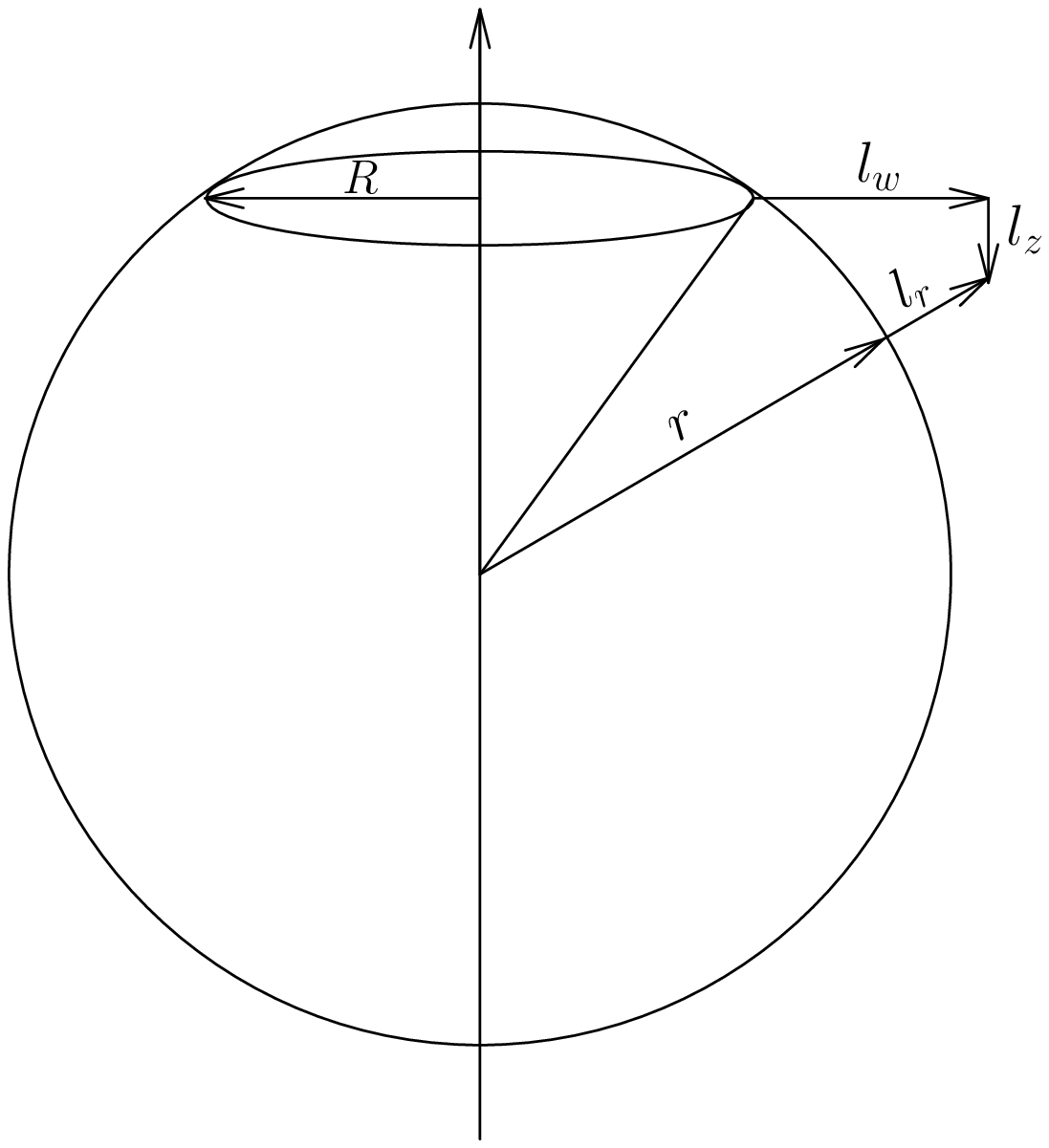}
\caption{Length scales of the displacement caused by the Tayler-Spruit instability.
        }
\label{fig:f1}
\end{figure}

In the form of the Acheson dispersion equation used by Spruit (1999), $l_{\rm h}$ actually stands for a distance from
the rotation axis, i.e. $l_{\rm h}\equiv l_w$ in the cylindrical polar coordinates $(w,\varphi,z)$, while
$l_r\equiv l_z$ is a displacement along the rotation axis. Let $R$ denote the radius of 
a magnetic ring that initially is concentric with the rotation axis and lies on a sphere of the radius $r$.
The Tayler-Spruit instability will cause a leading edge of the ring to move
a distance $l_w$ perpendicular and a distance $l_z$ parallel to the rotation axis (Fig.~\ref{fig:f1}).
As a result, a radial displacement of the leading edge will be
\bea
l_r = r\left(\sqrt{1+2\frac{Rl_w}{r^2}+\left(\frac{l_w}{r}\right)^2+
\left(\frac{l_z}{r}\right)^2-2\frac{l_z}{r}\sqrt{1-\left(\frac{R}{r}\right)^2}}-1\right).
\label{eq:sqrt}
\eea

Because Spruit's form of the dispersion relation is defined at the poles,
motions in the $w$ direction are assumed to
be perpendicular to gravity; this is not in general true for a spherical
geometry.  We can check on the maximum length scale
where this approximation breaks down as follows.  Assume that we start at
the rotation axis ($R=0$) and have $l_z = 0$. 
We consider that the assumptions in the
dispersion relation break down when the energy
loss from motions in the $w$ direction ($l_r^2N^2$) equals the energy gained
by the magnetic field ($r^2\omega_{\rm A}^2$).
Under these assumptions, equation (\ref{eq:sqrt}) can be re-written as
$$
\frac{l_r}{r} = \sqrt{1+\left(\frac{l_w}{r}\right)^2} - 1 = \frac{\omega_{\rm A}}{N}\ll 1,
$$
from which it immediately follows that $l_w < \sqrt{3(\omega_{\rm A}/N)}\,r\ll r$.
Indeed, in the deep solar core $N\approx N_\mu > N_T\sim 10^{-3}$, while
$\omega_{\rm A}\ll\Omega\sim 10^{-6}$, hence $l_w < 0.05\,r\ll r$.
For longer horizontal length scales
than this, the neglect of curvature terms in the
dispersion relation cannot be justified; motion that is assumed to gain
energy in the dispersion relation actually loses energy
in a spherical star.  This is consistent with Acheson's discussion of the
dispersion relation, where he notes
that it should be used only locally and not applied globally.  We therefore
conclude that the assumption $l_w \sim l_{\rm h}\sim r$ cannot
be obtained rigorously from the dispersion relation, and that other
considerations (such as energy and momentum conservation)
should be used instead to determine the maximum possible length
scales.  However, the relationship between the horizontal
and vertical length scales derived from the dispersion relation is valid
locally near the pole.

Another way of looking at this problem is to consider the consequences of 
Spruit's assumption that $l_w\sim l_{\rm h}\sim r$ in equation (\ref{eq:sqrt}).
If we take into account that $R$ cannot be less than $l_w$ (otherwise, the ring will
intersect the rotation axis which is not allowed in the Tayler-Spruit instability)  and that $l_z\leq l_w(\omega_{\rm A}/N)\ll l_w$
(this is a correct form of eq. \ref{eq:up2}) then we will find that $l_r\sim r$ as well. This is in contradiction with Spruit's 
estimate of $l_r\la r(\omega_{\rm A}/N)\ll r$. Equation (\ref{eq:sqrt}) shows
that $l_r$ can be much less than $r$ only if $l_w$ is much less than $r$ as well.
Thus, this simple analysis demonstrates that: {\it (i)} the assumption of $l_h\sim r$
is ruled out by the geometry of the problem, and {\it (ii)} even our revised diffusivities
obtained under the assumption that $l_{\rm h}\sim r\,(\omega_{\rm A}/\Omega)\ll r$ (see below)
may give overestimated values because, like Spruit, we did not discriminate between $l_{\rm h}$
and $l_w$ and between $l_r$ and $l_z$.

\subsection{The Coriolis Force}
\label{sec:coriolis}

Our revised magnetic diffusion coefficient and viscosity differ from those obtained by
\cite{s99} only because we have used a different
(reduced by the factor of $\omega_{\rm A}/\Omega$) upper limits for the horizontal and radial
wavelengths of the unstable fluid displacement in (\ref{eq:up}).
In this section, we show that our
choice of $l_r$ is physically more consistent than that made by Spruit 
as we take into account a reduction
of the horizontal wavelength $l_{\rm h}$ by the Coriolis force.

To prove that his heuristic derivations yielded correct upper and lower limits for $l_r$,
Spruit analysed the Acheson (1978) dispersion relation using a few simplifications.
The original Acheson equation includes all relevant physical
processes in an electrically conducting spherical body rotating in the presence of
a toroidal magnetic field and its own gravitational field. 
The analysis of the dispersion relation in the limit of $\Omega\gg\omega_{\rm A}$
has led Spruit to the conclusion that in the case of $K=0$, which is applicable
when $N_\mu\gg N_T$, sufficient and sufficiently accurate necessary conditions
for oscillatory modes of the Tayler-Spruit instability are both met if    
\bea
l_r^2\la l_{\rm h}^2\,\frac{\omega_{\rm A}^2}{N^2},
\label{eq:up2}
\eea
and
\bea
l_r^2\ga\frac{\eta}{(\omega_{\rm A}^2/\Omega)}.
\label{eq:low2}
\eea
Whereas the second condition is already identical with the heuristic inequality (\ref{eq:low}),
the first condition would coincide with the inequality (\ref{eq:up}) only if 
the horizontal wavelength $l_{\rm h} = r$
(assuming, of course, that $\omega = \omega_{\rm A}$ in (\ref{eq:up}), as Spruit did).

The important point we want to emphasize here is that the condition (\ref{eq:up2}),
derived from the Acheson equation, does not give a direct estimate of
the radial length scale $l_r$. Instead, it says that the maximum possible
value of $l_r$ is proportional to $l_{\rm h}$
with a small coefficient $\omega_{\rm A}/N$. In order to get a final estimate for $l_r$,
we have to make a physically motivated choice of $l_{\rm h}$. Spruit chose
``the longest possible horizontal wavelength'' $l_{\rm h}\sim r$. But his choice is based
on a purely geometric constraint which can be applied only if there are no other
constraints that bound the unstable displacement along the horizontal surface.
We argue that Spruit's choice is only valid in the absence of rotation
(see, however, \sect{sec:geometry}).

In a rotating star (here, we assume the uniform rotation, like Spruit did when he analysed the Acheson
equation), the Coriolis force $2\mathbf{v}\times\mathbf{\Omega}$
can considerably reduce the horizontal length scale of fluid motions.
To demonstrate this, we assume that the leading edge of the ring
has the initial velocity ${\mathbf v} = \{v_{\rm A},0\}$ directed off the rotation axis
along the X axis (Fig.~\ref{fig:f3}). Here, $v_{\rm A} = r\,\omega_{\rm A}$ is the Alfv\'{e}n
velocity, and the coordinate plane XOY is chosen to be perpendicular to the rotation axis.
Our assumption is consistent with the development of the Tayler instability.
Having solved the momentum equation, we find that the motion of the ring under the action of
the Coriolis force is described by the following equations:
\bea
x & = & R\cos\theta + r\,\frac{\omega_{\rm A}}{2\Omega}\,\sin(2\Omega t), \nonumber \\
y & = & R\sin\theta + r\,\frac{\omega_{\rm A}}{2\Omega}\,\left[\,\cos(2\Omega t)-1\right],
\label{eq:motion}
\eea
where $R$ is the radius of the ring, and the angle $0\leq\theta < 2\pi$ specifies a position
of a point on the ring. In the limit of $\Omega\rightarrow 0$ (no rotation), these equations
are transformed into $x\rightarrow R\cos\theta + r\omega_{\rm A}t$ and $y\rightarrow R\sin\theta$,
i.e. they describe a displacement of the ring with the constant velocity $v_{\rm A}$ along the X axis.
However, when $\Omega\neq 0$ the Coriolis force will deflect the ring from this rectilinear
motion. For example, for a particular choice of $r=10$, $R=2$ (in arbitrary units), and
the ratio $\omega_{\rm A}/\Omega = 0.1$, consecutive positions (for the phases
$2\Omega t = 0,\,\pi/2,\,\pi$, and $3\pi/2$) of the ring are plotted
in Fig.~\ref{fig:f3}. From equations (\ref{eq:motion}) it follows that, in the case of $\omega_{\rm A}\ll\Omega$, 
the Coriolis force restricts the ring motion within a distance of $l_{\rm h}\sim r(\omega_{\rm A}/\Omega)$
around its original location. In the limit of $\omega_{\rm A}/\Omega\rightarrow 0$,
equations (\ref{eq:motion}) are reduced to $x\rightarrow R\cos\theta$ and $y\rightarrow R\sin\theta$,
i.e. in this extreme case $l_{\rm h}\rightarrow 0$, and the ring cannot move horizontally at all.

It is interesting that a similar restriction imposed by the Coriolis force has been
studied by \cite{chg87} in relation to the axisymmetric rise of a magnetic ring in the solar convective zone
caused by the buoyancy instability. Like in our case, although the Acheson equation
gives the same reduced growth rate $\omega\sim\omega_{\rm A}^2/\Omega$ for the buoyancy
instability, it does not predict how the magnetic ring is deflected by the Coriolis force
on the large length scales. To find this out, one needs to consider the dynamics of large scale
displacements.

What happens if $\Omega$ is not held constant during the non-axisymmetric oscillations?
For a ring of size $R$ perturbed by a distance $l_w$ from the rotation axis, angular momentum
conservation implies
$$
\Omega = \Omega_0\frac{R^2}{R^2+l_w^2}.
$$
There will be an oscillating angular velocity gradient
produced by this with a characteristic vertical length scale $l_z$ of magnitude
$$
\frac{d\Omega}{dz} = \frac{\Omega_0 - \Omega_0\frac{R^2}{R^2+l_w^2}}{l_z} =
\frac{\Omega_0}{l_z}\frac{l_w^2}{R^2+l_w^2},
$$
hence
$$
q = \frac{z}{\Omega_0}\frac{d\Omega}{dz} = \frac{zl_w^2}{l_z(R^2+l_w^2)}.
$$
Since $l_w/l_z \gg 1$, this implies a large $q$ unless $zl_w/(R^2+l_w^2)\ll 1$,
which corresponds to small motions of large rings. A back reaction would therefore be produced, damping the motions, unless
$\Omega\sim$\,constant were maintained, hence this is a reasonable assumption.

Furthermore, taking into account the fact that the Coriolis force induces the azimuthal motion,
the assumption of the star's uniform rotation made by Spruit seems inconsistent. 
The natural question arises ---
what causes the corotation of the fluid element with its surroundings
when it moves off the rotation axis along the horizontal
surface or, in other words, what force compensates the Coriolis force? 
In Spruit's derivation, this cannot be a viscous friction because viscosity is neglected in the version of the Acheson
equation used by him. A relevant example from the Earth's atmosphere science is the
geostrophic wind. This wind can flow along a meridian only thanks to the presence of a horizontal
pressure gradient that balances the side action of the Coriolis force. In stars, there is no
such gradient on the equipotential surfaces. Rotation-driven shear instabilities in stars
may result in a strongly anisotropic turbulent viscosity with its horizontal component
strongly dominating over the vertical component (\citealt{z92}). So, in fact, it is rather
the horizontal viscous friction, not taken into account by Spruit, that may convert the azimuthal kinetic energy of
the fluid element into the kinetic energy of the horizontal turbulence.

\subsection{Support from the Magnetic Induction Equations}
\label{sec:induction}

Our result is derived by limiting the horizontal length scale in the presence of rotation.
The Coriolis force bounds $l_{\rm h}$ (or, to be more precise, $l_w$) by the limit of order
$r(\omega_{\rm A}/\Omega)$, which ensures the angular momentum conservation. In this section, we
demonstrate that our reduced upper limit on $l_{\rm h}$, combined with the limit (\ref{eq:up2}) on the ratio of
the radial and horizontal length scales obtained by Spruit (1999), is consistent
with the requirement that the excess kinetic energy in differential rotation --- the only source of
energy in the problem --- always exceeds the work needed to be done against the buoyancy during the radial
displacement.

Assuming that all of the requirements (\ref{eq:req}) are satisfied, we can write
\bea
B_\varphi^2\approx\Omega^2q^2B_r^2\tau^2,
\label{eq:bphi2}
\eea
where $\tau\sim\omega^{-1}=\Omega/\omega_{\rm A}^2$. On the other hand,
\bea
B_r^2\approx\frac{l_r^2}{r^2}B_\varphi^2=4\pi\rho\, l_r^2\omega_{\rm A}^2.
\label{eq:br2}
\eea
In the last equality, we used the definition of $\omega_{\rm A}$ (eq. \ref{eq:s1}).
Combining equations (\ref{eq:bphi2}\,--\,\ref{eq:br2}), we obtain
\bea
\frac{B_\varphi^2}{8\pi}= \frac{1}{2}\rho\, r^2\omega_{\rm A}^2 \approx \frac{1}{2}\rho\,\Omega^2q^2l_r^2\left(\frac{\Omega}{\omega_{\rm A}}\right)^2.
\label{eq:final}
\eea
Earlier we found that $N=\Omega q$. There is a natural energetic argument for this (Maeder 2006, private communication).
The product $\frac{1}{2}\rho\,\Omega^2q^2l_r^2$ measures an excess of kinetic energy (per unit volume) in the differential
rotation between spherical shells separated by a radial distance $l_r$. This is the only
source of energy available to do the work $\frac{1}{2}\rho\, l_r^2N^2$ against the Archimedes force to produce
mixing on the length scale $l_r$ in the radial direction. Without this mixing, no radial
magnetic field will be generated by the magnetic induction. This conclusion is especially
evident when we consider a deep radiative core of the star where $N^2\approx N_\mu^2$.
Here, mixing itself cannot affect the value of $N^2$. If $\Omega^2q^2$ is less
than $N_\mu^2$ then the differential rotation simply does not possess a sufficient amount of
the excess kinetic energy to overcome the stable thermal stratification. It does not matter
what sophisticated mechanisms we may invent to transform this energy into other forms.
In the end, it will still have to exceed the work $\frac{1}{2}\rho\, l_r^2N_\mu^2$ and, by the energy conservation
law, it will still be equal to $\frac{1}{2}\rho\,\Omega^2q^2l_r^2$ at most.
This is why our revised magnetic diffusivities vanish whenever $\Omega^2q^2 < N_\mu^2$.
Substituting the inequality $\Omega^2q^2\geq N^2$ into (\ref{eq:final}), we immediately
get our estimate of $l_r\leq r\omega_{\rm A}^2/N\Omega$.

It should be noted that eq. (\ref{eq:final}) is in concordance with our conclusion
that the Coriolis force reduces the horizontal length scale by the factor of $\omega_{\rm A}/\Omega$, provided
that $\frac{1}{2}\rho\,\Omega^2q^2l_r^2 = \frac{1}{2}\rho\, l_r^2N^2$.
Indeed, under this assumption eq. (\ref{eq:final}) can be written as
\bea
\frac{1}{2}\rho\,\Omega^2q^2l_r^2 = \frac{1}{2}\rho \,l_r^2N^2 =\frac{1}{2}\rho\, r^2\omega_{\rm A}^2\left(\frac{\omega_{\rm A}}{\Omega}\right)^2.
\label{eq:ind}
\eea
After the subsitution of $l_r=l_{\rm h}(\omega_{\rm A}/N)$ into the mid term
and the comparison of it with the last term,
we will immediately find that $l_{\rm h}=r(\omega_{\rm A}/\Omega)$. This means that
the latter relation is actually equivalent (in the mathematical sense) to the energy balance equation
$\frac{1}{2}\rho\,\Omega^2q^2l_r^2 = \frac{1}{2}\rho\, l_r^2N^2$.
Equation (\ref{eq:ind}) shows that only a small fraction of the energy of the toroidal magnetic field is used
to do the work against the buoyancy.
On the other hand, if we had followed Spruit's assumption
that $\frac{1}{2}\rho r^2\omega_{\rm A}^2 = \frac{1}{2}\rho l_r^2N^2$ then
we would have found from eq. (\ref{eq:final}) that $\frac{1}{2}\rho\Omega^2q^2l_r^2\ll \frac{1}{2}\rho l_r^2N^2$. In that case,
some extra energy source, besides the differential rotation, would have been needed to drive
the Tayler-Spruit dynamo.

\subsection{A Common Sense Argument}
\label{sec:common}

Equations (\ref{eq:oldetae}) and (\ref{eq:nue2}) show that, in the limit of $N_T\gg N_\mu$,
Spruit's magnetic viscosity does not depend on the shear $q$ at all.
This is difficult to understand conceptually because it is the differential rotation that
makes the whole Tayler-Spruit mechanism work. In our revised prescription,
the dependence of diffusivities on $q$ holds in all regimes.

\section{Comparison with Observations}
\label{sec:sun}

The original Spruit (2002) diffusion coefficients for angular momentum transport were 
both large and independent of the angular velocity gradient in the absence of $\mu$ gradients, 
and mildly sensitive to $\mu$ gradients.  The Maeder \& Meynet (2004) coefficients have been derived 
in a more self-consistent fashion, but exhibit the same global properties.  
Our revised prescription has fundamentally different dependencies on both $\Omega$ and $\mu$ gradients, 
which has an important impact on the behavior of stellar models.  One important test of 
the impact of the Tayler-Spruit mechanism is the inferred solar rotation curve, 
which we estimate below.  The older diffusion coefficient estimates were consistent 
with the solar core rotation (Eggenberger et al. 2005), and the revised coefficients 
predict core rotation too high to be compatible with seismology.

However, there are two other important observational constraints on angular momentum transport 
in low mass stars that are also important: the spindown of young cluster stars and 
the preservation of sufficient angular momentum in red giant cores to explain 
the rapid rotation of horizontal branch stars.  In both cases the observational data imply 
that the timescale for efficient core-envelope angular momentum transport is substantially 
less than the age of the Sun, but comparable to the lifetimes of the relevant 
phases (of order 20-100 Myr); i.e. solid-body rotation enforced at all times is a poor fit 
(\citealt{kea95,kea97,a98,sea00}).
Eggenberger et al. (2005) found that their 
diffusion coefficients predicted essentially solid-body rotation at all times, 
which would not be consistent with young open cluster data.  We defer a detailed discussion of 
the spindown properties of the revised prescription to a paper in preparation, 
but we will present evidence here that the predicted timescale for core-envelope coupling is 
longer in the new prescription than in the older one.  Adopting the revised coefficients 
therefore improves the consistency between theoretical models and the data, 
by permitting the observationally required deviations from solid-body rotation.  
Combining these two sets of constraints, we therefore conclude that some other 
mechanism is required to explain the slow rotation of the solar core, either another 
magnetic instability or wave-driven momentum transport; however, inclusion of 
the Tayler-Spruit instability may be a promising explanation for the timescale of core-envelope coupling.

As a test, we compare our revised transport coefficients for the Tayler-Spruit dynamo 
to previous results for a solar model. We employ the stellar evolution code described 
by \cite{dea06}; gravitational settling was not included.  We used the \cite{gn93} mixture of 
heavy elements and calibrated our model to reproduce the solar luminosity 
($3.85\times 10^{33}$\,erg\,s$^{-1}$) and radius ($6.96\times 10^{10}$\,cm) at 
the solar age of 4.57 Gyr; this procedure yielded $Y = 0.273$, $Z = 0.018$, and a mixing length $\alpha$ of 1.75.  

For angular momentum evolution, it is important to specify the initial conditions and boundary conditions.  
Our computations start on the pre-MS deuterium birthline of \cite{ps91}
with an initial rotation period of 8 days, close to the mean value for classical T Tauri stars.  
Solid-body rotation was enforced in convective regions at all times.  
No interaction between the protostar and disk was included, implying that 
the model considered corresponds to a rapidly rotating young open cluster star.  
We adopt the angular momentum loss prescription of \cite{kea97}:
$$
\dot{J}_{\rm tot} = -K_{\rm w}\sqrt{\frac{R/R_\odot}{M/M_\odot}}
\,\min\{\Omega_{\rm s}^3,\,\Omega_{\rm s}\Omega_{\rm crit}^2\}.
$$
where $R$ and $\Omega_{\rm s}$ are the star's surface radius and angular velocity.  
$\Omega_{\rm crit}$ is the critical velocity at which angular momentum loss from the magnetized stellar wind gets saturated.
Following \cite{kea97}, we adopt $\Omega_{\rm crit} = 10\,\Omega_\odot$, where $\Omega_\odot\approx 2.87\times 10^{-6}$\,rad\,s$^{-1}$.
The parameter $K_{\rm w}$ is calibrated requiring that our evolved models have $\Omega_{\rm s}=\Omega_\odot$
at the solar age ($K_{\rm w}=1.1\times 10^{47}$\,cm$^2$\,g\,s$^{-2}$ for our unmixed model). 

With these basic ingredients we consider three angular momentum transport cases: 
a model without transport in the radiative interior; a full evolutionary model using 
the \cite{mm04} coefficients; and solutions with the revised coefficients employing static 
solar mass and abundance models at different ages.  

For comparison with the rotation profile obtained by \cite{eea05}, we used
the old relations (\ref{eq:mm04}\,--\,\ref{eq:4thcoeff}), together with
the equation of the transport of angular momentum by magnetic viscosity:
\bea
\frac{\partial}{\partial t}(r^2\Omega) = \frac{1}{r^2\rho}\frac{\partial}{\partial r}
\left(r^4\rho\,\nu_{\rm e}\frac{\partial\Omega}{\partial r}\right).
\label{eq:am}
\eea

In our revised diffusion coefficient estimates the quantities
$\eta_{\rm e}$ and $\nu_{\rm e}$ depend strongly on both $\Omega$ and $q$, and 
there is a steep reduction in the efficiency of angular momentum transport 
in the presence of $\mu$ gradients. These features make it numerically more challenging 
to calculate them and $q$ with a reasonable accuracy.  However, we can estimate a lower bound 
to the solar rotation curve reliably under the following set of assumptions.  
First, we assume that the timescale for angular momentum transport is less than the solar age in the absence of $\mu$ gradients.  
However, there is a threshold value of the $\mu$-gradient above which the new coefficients vanish. 
Assuming that $\eta_{\rm e} = 0$ in (\ref{eq:etae}), the threshold $\mu$-gradient can be estimated
from the following equation:
\bea
N_\mu^2\equiv g\varphi\left|\frac{\partial\ln\mu}{\partial r}\right| = \Omega^2\,q^2,
\label{eq:thres}
\eea
where $g$ is the local gravity.
Although formally $\eta_{\rm e}$ cannot be less than the magnetic diffusivity $\eta$, 
the ratio $\eta/K$ is so small that equation (\ref{eq:thres}) gives a very accurate estimate of 
the threshold $\mu$-gradient.  We integrated this condition from the surface to the point 
where the local rotation would be higher than the no-transport case for three different 
snapshots of the solar composition profile (at 150 Myr, 600 Myr, and 4.57 Gyr).  

\begin{figure}
\plotone{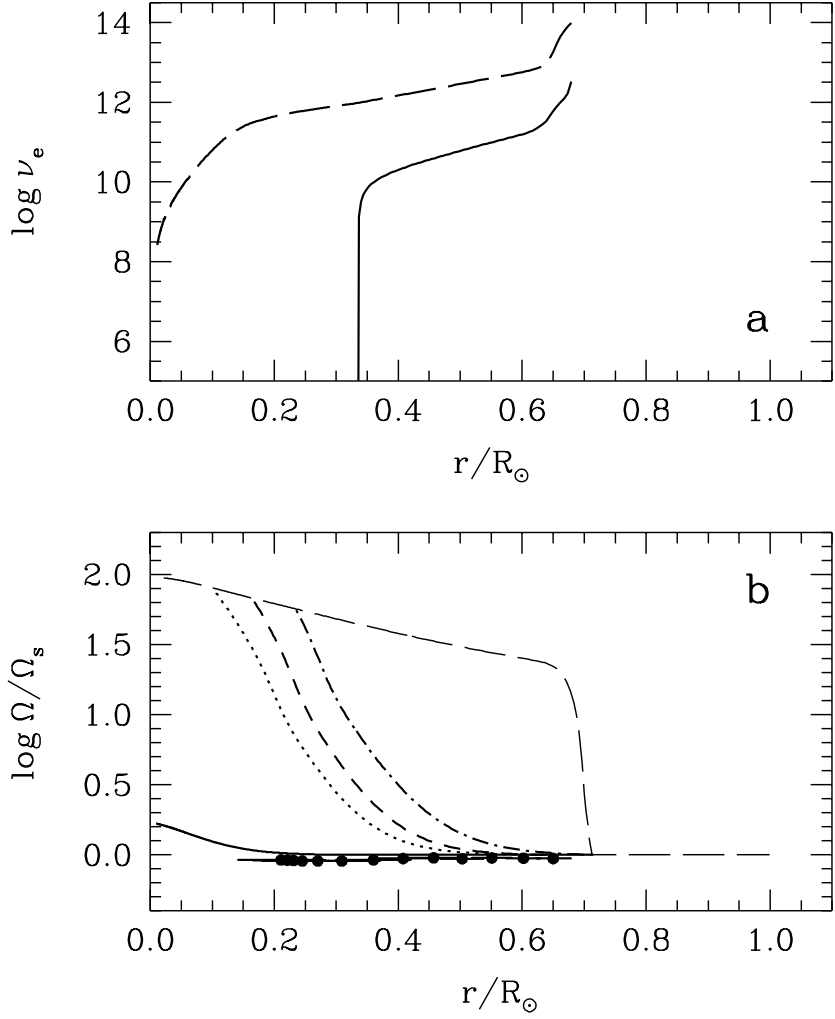}
\caption{Panel a: the old ({\it long-dashed curve}) and new ({\it solid curve}) 
         magnetic viscosities for the no-transport rotation profile ({\it long-dashed curve} in
         panel b).
         Panel b: 
         {\it solid curve} --- rotation profile from our full evolutionary computations obtained
         as a solution of eq. (\ref{eq:am}) using the \cite{mm04} prescription for $\nu_{\rm e}$;
         {\it dotted, short-dashed} and {\it dot-dashed
         curves} --- rotation profiles for our revised $\nu_{\rm e}$ estimated with equation (\ref{eq:thres})
         using $N_\mu^2$ from models of ages $1.5\times 10^8$, $6\times 10^8$, and $4.57\times 10^9$ yr,
         respectively; {\it dots} are helioseismic data from \cite{cea03}.
        }
\label{fig:f3}
\end{figure}

Our results are summarized in Fig.~\ref{fig:f3}.  In the bottom panel, we compare 
the normalized solar rotation as a function of radius obtained from 
the helioseismic data of \cite{cea03} ({\it dots} with errorbars) to a model 
without angular momentum transport (the long-dashed line), a full evolutionary model 
using the \cite{mm04} prescription (solid line), and rotation curves derived 
from static models with $\mu$ profiles appropriate for solar models at ages of 150 Myr, 600 Myr, and 4.57 Gyr (dotted,
 short-dashed and dot-dashed lines, respectively).  In the top panel, 
we compare the magnetic viscosities that we would have obtained for the no-transport 
rotation curve using our prescription (the solid line) and the \cite{mm04} 
prescription (the dashed line).  The results in the top panel indicate that magnetic 
angular momentum transport would be highly effective for stellar envelopes for both 
the new and old formulations of the Tayler-Spruit instability.  The new diffusion coefficients, 
however, are 2-3 orders of magnitude lower than the previous values outside the solar core, 
and vanish completely in the deep solar interior.

When compared with the solar data, the marginal stability cases for magnetic transport 
with our diffusion coefficients are clearly well above that permitted by helioseismology.  
This effect is significant even for composition gradients appropriate for the early MS 
(the 150 Myr and 600 Myr curves).  This mechanism is therefore not 
responsible for the slow rotation of the solar core.  However, the reduction 
in the inferred envelope diffusion coefficients may bring them into a range compatible 
with the spindown of young cluster stars.  We conclude that, at least for 
the evolution of low mass stars, inclusion of this magnetic transport mechanism may have 
an interesting impact on some observationally important features of models including rotation, 
but that some other effect is still missing in our understanding of the internal solar rotation.

\acknowledgements
We are grateful to Grant Newsham, Don Terndrup and two anonymous referees for their
comments and suggestions that have served to improve this paper. We also thank people
who commented on the first version of our paper in astro-ph/0604045. Tony Piro brought our
attention to subtle flaws in our original argument based on the dispersion relation.
Andre Maeder mentioned the relationship between the energy of mixing and the energy of
differential rotation, and Henk Spruit provided helpful insight into the dispersion relation.
We acknowledge support from the NASA Grant No. NNG05 GG20G.



\clearpage



\clearpage
\begin{figure}
\plotone{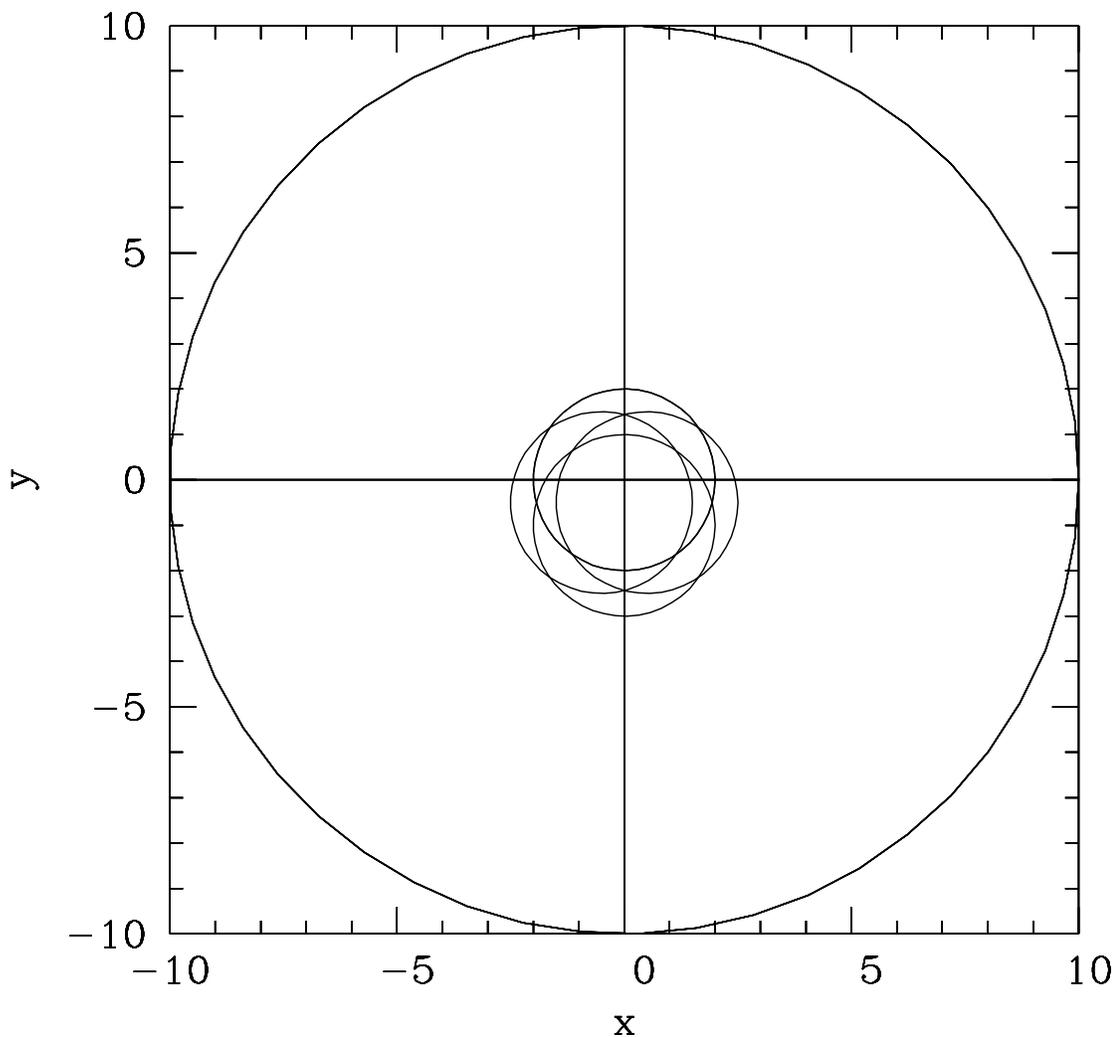}
\caption{An illustration of the motion of a magnetic ring of the radius $r=2$ (in arbitrary 
         units) constrained by the Coriolis force (eqs. \ref{eq:motion}). 
         The large circle is the equator of the sphere of the radius $r=10$ inside the star.
         The coordinate plane XOY is perpendicular to the rotation axis. The smaller circles
         show consecutive positions of the ring 
         at the phases $2\Omega t = 0,\,\pi/2,\,\pi$, and $3\pi/2$.
         The ring starts its motion from the pole with
         the velocity ${\mathbf v}=\{v_{\rm A},\,0\}$, where $v_{\rm A}=r\,\omega_{\rm A}$.
         The plotted case corresponds to the ratio $\omega_{\rm A}/\Omega = 0.1$.
         }
\label{fig:f2}
\end{figure}



\clearpage


\end{document}